\documentclass[10pt,final,twocolumn,journal]{IEEEtran}
\usepackage[noadjust]{cite}
\usepackage{cite}
\usepackage{graphicx}
\usepackage{caption}
\usepackage{subcaption}
\usepackage{authblk}
\usepackage{xcolor}
\usepackage{amsmath}
\usepackage{array}
\usepackage[font=small,labelfont=bf]{caption}

\begin{document}
\title{Impact of Dimensionality on PN Junctions}
\vspace{-3.3\baselineskip}
\author{Hesameddin Ilatikhameneh, Tarek Ameen, Fan Chen, Harshad Sahasrabudhe, Gerhard Klimeck, Rajib Rahman \vspace{-6.5ex}
\thanks{This work was supported in part by the Center for Low Energy Systems Technology (LEAST), one of six centers of STARnet, a Semiconductor Research Corporation program sponsored by MARCO and DARPA.}
\thanks{The authors are with the Department of Electrical and Computer Engineering, Purdue University, West Lafayette, IN, 47907 USA e-mail: hesam.ilati2@gmail.com.}
}
\maketitle
\setlength{\textfloatsep}{12pt} 
\setlength{\belowdisplayskip}{1.6pt} 
\setlength{\belowdisplayshortskip}{1.6pt}
\setlength{\abovedisplayskip}{1.6pt} 
\setlength{\abovedisplayshortskip}{1.6pt}
\setlength{\belowcaptionskip}{-12pt}
\vspace{-1.0\baselineskip}
\begin{abstract}
Low dimensional material systems provide a unique set of properties useful for solid-state devices. The building block of these devices is the PN junction. In this work, we present a dramatic difference in the electrostatics of PN junctions in lower dimensional systems, as against the well understood three dimensional systems. Reducing the dimensionality increases the fringing fields and depletion width significantly. We propose a novel method to derive analytic equations in 2D and 1D that considers the impact of neutral regions. The analytical results show an excellent match with both the experimental measurements and numerical simulations. The square root dependence of the depletion width on the ratio of dielectric constant and doping in 3D changes to a linear and exponential dependence for 2D and 1D respectively. This higher sensitivity of 1D PN junctions to its control parameters can be used towards new sensors. 
\end{abstract}
\vspace{-0.5\baselineskip}
\vspace{-1.0\baselineskip}
\section{Introduction}
Low dimensional materials have become the building blocks of nanotechnology. The fundamental properties of 1D and 2D materials offer a new toolbox for electronic \cite{Geim}, thermoelectric \cite{Dres}, and optoelectronic  \cite{MoS2_photo} applications; like the high mobility graphene, carbon nanotube, and III-V nanowires for transistors \cite{FrankGR, CNT, Fan1, Fiori2, Fan2, MoS2, Mehdi}, transition metal dichalcogenides and phosphorene for tunnel transistors \cite{Hesam1, Sarkar, Hesam3, Fiori_Nature, Hesam2, Tarek1, Hesam4}, high ZT nanoribbon and nanowire thermoelectrics \cite{Thermo, Si_Thermo, Si_Thermo2} and 2D photodetectors \cite{MoS2_photo2, Tao}. Almost all electronic and optical devices are based on PN junctions. Hence, the knowledge of the electrostatics of PN junctions is critical in predictive device design. For example, the tunneling current in tunnel diodes \cite{Kane1, SB1, Analytic2} and tunnel FETs \cite{App1, Ionescu, Sub10, Analytic1} depends exponentially on the depletion width. Low dimensional material systems, however, present significant differences compared to their 3D counterparts, as detailed in the next section. The rapid progress in doping thin flakes of 2D materials and 1D nanowires over the last two decades \cite{Doping1, Doping2, Doping3} calls for a thorough analytical understanding of the low dimensional PN junctions.

In this work, a novel approach is proposed for analytic modeling of PN junctions and the equations governing the potential profiles and depletion width of 3D, 2D and 1D systems are presented. Experiments on PN junctions demonstrate a dramatic impact of dimensionality on the electrostatics \cite{Exp2D, Exp1D, Changxi}. Fig. 1 shows the simulated potential profile of PN junctions, in different dimensions, with the same doping of 5e19 cm$^{-3}$ using finite-element method. It is clear that the well known 3D electrostatics does not apply to low dimensional junctions. 
\begin{figure}[!t]
        \centering
        \begin{subfigure}[b]{0.42\textwidth}
               \includegraphics[width=\textwidth]{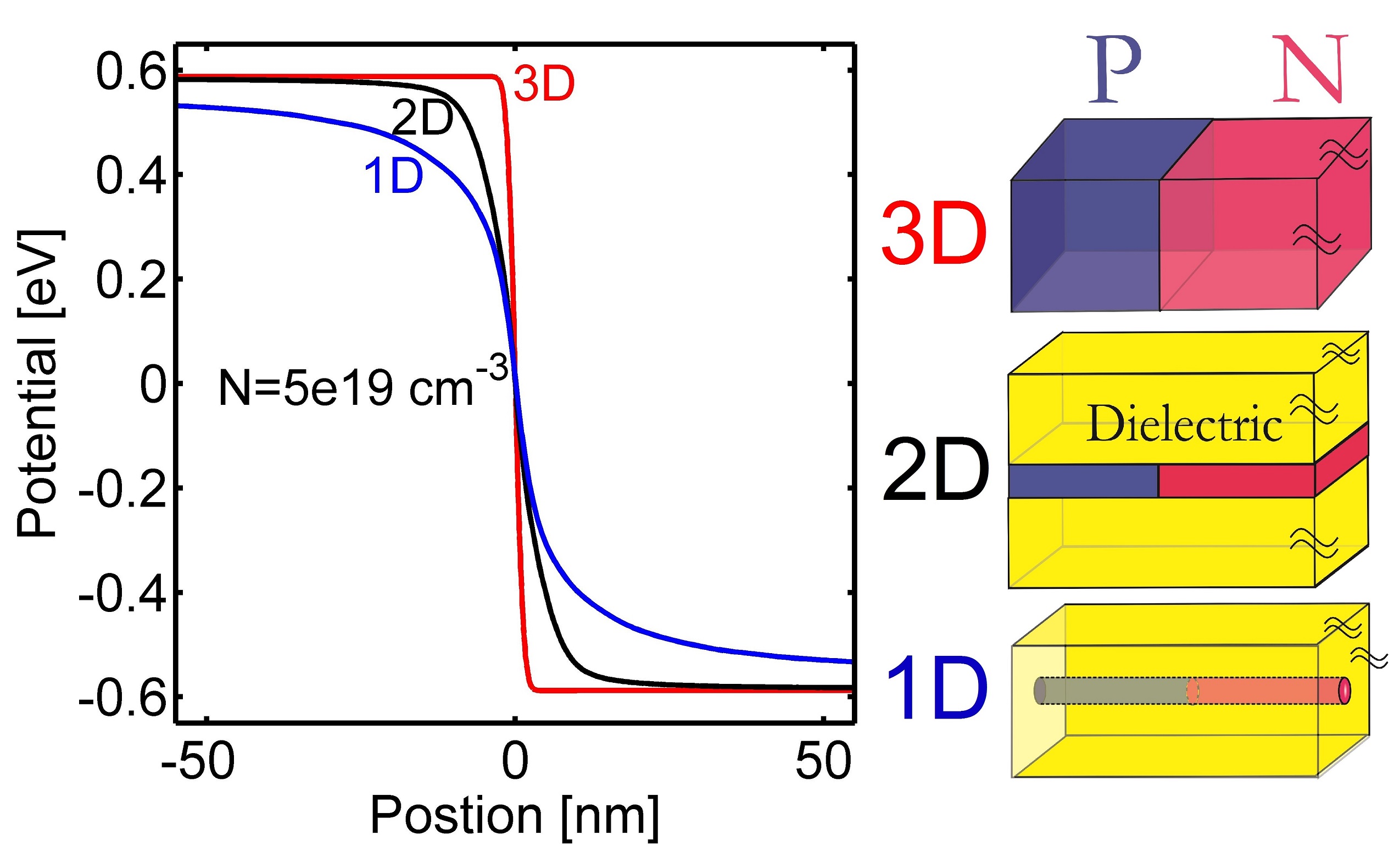}
                \label{fig:Same_EoT}
        \end{subfigure}%
        \vspace{-1.5\baselineskip}
        \caption{Potential profile of PN junctions in 3D, 2D, and 1D. }\label{fig:Fig1}
\end{figure}
The previous analytic efforts neglect the impact of neutral regions. The high carrier densities in neutral regions screen the electric field. Ignoring these regions in analytic models can lead to non-monotonic potential profiles \cite{Khor1, Khor2}, in contradiction with simulation results. 

Table I shows the approximate equations for depletion widths ($W_D$) of PN junctions in different dimensions as obtained in this work. These approximations are valid when the thickness of PN junction is much smaller than $W_D$. For the sake of simplicity, it is assumed that the P and N doping levels equal $N$ and the dielectric constants of channel and surrounding dielectric equal $\epsilon$.  Interestingly, the square root dependence of $W_D$ on $\epsilon / N$ in 3D, changes to a linear and exponential dependence for 2D and 1D, respectively. Such an exponential dependence leads to high sensitivity of 1D PN junction to its control parameters, which can be used towards new sensors. For example, a small change in the biasing of 1D PN junction affects $W_D$ significantly through $\Delta V$ (i.e. $V_{bi}-V_{a}$ where $V_{bi}$ is the built-in potential and $V_{a}$ is applied bias). Such a junction under illumination can result in a dramatic output current response.    
\begin{table}[!b]
        \centering
\vspace{0.7\baselineskip}
\captionof{table}{Depletion widths of 1D, 2D, 3D PN junctions. $R$ and $T$ are nanowire radius and flake thickness in 1D and 2D, respectively.}             
\resizebox{0.4\textwidth}{!}{
\begin{tabular}{lr} 
$W_D$ & Dimension \\
\hline
$ 2 \sqrt{ \frac{\Delta V \epsilon}{q N} } $ & 3D  \\
$\frac{\pi \epsilon \Delta V}{ln(4) ~ q N T} $ & 2D \\
$ R ~ exp \left(\frac{2 \epsilon \Delta V}{q N R^2} - \frac{1}{2}\right)$ & 1D 
\end{tabular} }
\end{table} 
A novel approach to obtain the analytic equations, including the impact of neutral regions, is introduced first. Then the results of analytical model are compared with experimental measurements and numerical simulations. Finite element method is used to solve the Poisson equation self-consistently with the drift-diffusion using NEMO5 \cite{N5, N5_2}.  
\vspace{-0.8\baselineskip}
\section{Discussion and Results} 
In this section, the proper method to consider the impact of high carrier density in neutral regions is introduced. To demonstrate the new approach, we start with the well known 3D case. The depletion and neutral regions are shown in Fig. 2a. Considering only the charge in the depletion region wrongfully leads to a nonzero field at the boundaries of depletion region \cite{Khor1, Khor2}. Free carriers in the neutral regions play a very important role in screening the electric field. This effect can be captured by introducing image charges. Fig. 2b shows the equivalent structure where neutral regions are replaced by image charges. Fig. 2c shows the uncompensated charges which contribute to $E_{net}$, the net electric field at position x due to all the charges. Notice that the field due to charges from 0 to $x$ compensates the ones from $x$ to $2x$. Moreover there is no field due to charges positioned after $W_D/2 + x$ and before $-W_D/2 + x$ since these two regions induce opposite fields which cancel each other. This is because the net negative charge in region 1 is the same as the one in region 2 and since the electric fields are proportional to net charge in 3D, the fields due to the two regions cancel each other completely. As a result, $E_{net}$ is the field due to a block of negative charge of length $W_D/2 - x$ at a distance of $x$ on the left and a similar positive one on the right. Since these two blocks of charges produce same electric field, it is sufficient to calculate the field due to one of them as shown in Fig. 2d:    
\begin{equation}
\label{eq:opt1}
E_{net}^{3D}(x) = \frac{qN}{\epsilon} \left(x- \frac{W_D}{2} \right),~ 0 < x < \frac{W_D}{2}
\end{equation}
\begin{figure}[!t]
        \centering
        \begin{subfigure}[t]{0.4\textwidth}
               \includegraphics[width=\textwidth]{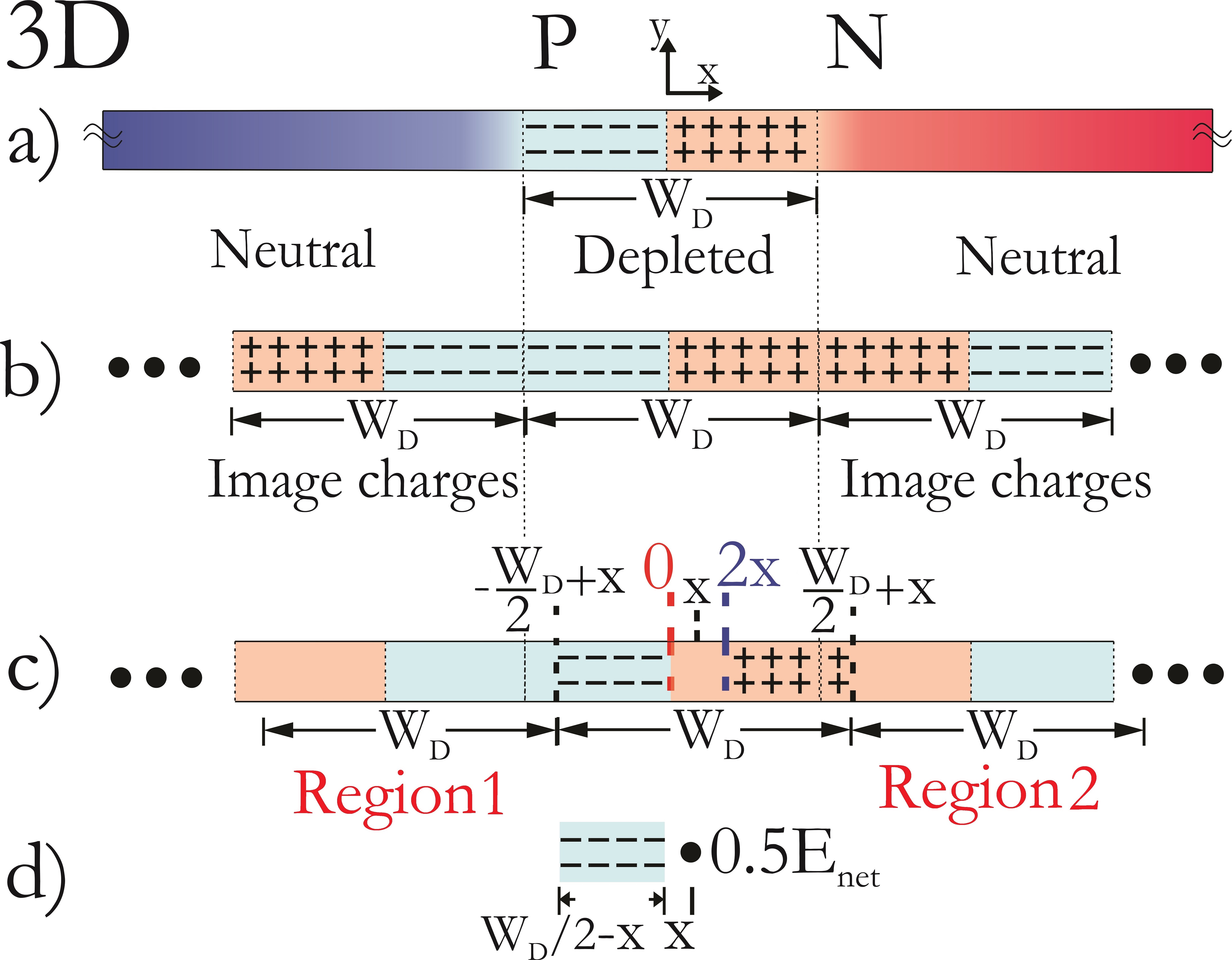}
        \end{subfigure}%
        \caption{Illustration of the analysis method in a 3D PN junction: a) netural and depletion regions, b) image charges, c) uncompensated charges, and d) net electric field.}\label{fig:Fig2}
\end{figure}
Integrating electric field gives the potential. Considering the reference potential to be the value at $x$=0 results in:
\begin{equation}
\label{eq:V3D}
V^{3D}(x) = - \int_0^x E(x') dx' = \frac{qN}{2 \epsilon} x \left( W_D - x \right)
\end{equation}
As expected, the potential profile has a parabolic form in the depleted region. From (\ref{eq:V3D}), $W_D$ can be derived: $W_D^{3D} = 2 \sqrt{ \frac{\Delta V \epsilon}{q N} }$.
%
Hence, we recover the well known electrostatics of bulk PN junctions.
The same procedure can be followed for 2D junctions to find the potential profile as shown in Fig. 3. However, there is a difference compared to the 3D; the net electric field due to charges after $W_D/2 + x$  and before $-W_D/2 + x$ is not zero since, unlike in 3D, the net field is not proportional to net charge in 2D and 1D junctions. These regions behave similar to electric dipoles with a finite net field. To the first order, the charged regions are extended till adjacent compensated region. Fig. 3c shows the uncompensated charges for a 2D junction. The impact of charges further away are discussed in the last section. Finally, the net electric field at a distance x is obtained from a 2D block of charge with length $W_D - 2x$ and thickness $T$ and multiplied with a factor of 2 as shown in Fig. 3d:
\begin{figure}[!t]
        \centering
        \begin{subfigure}[b]{0.4\textwidth}
               \includegraphics[width=\textwidth]{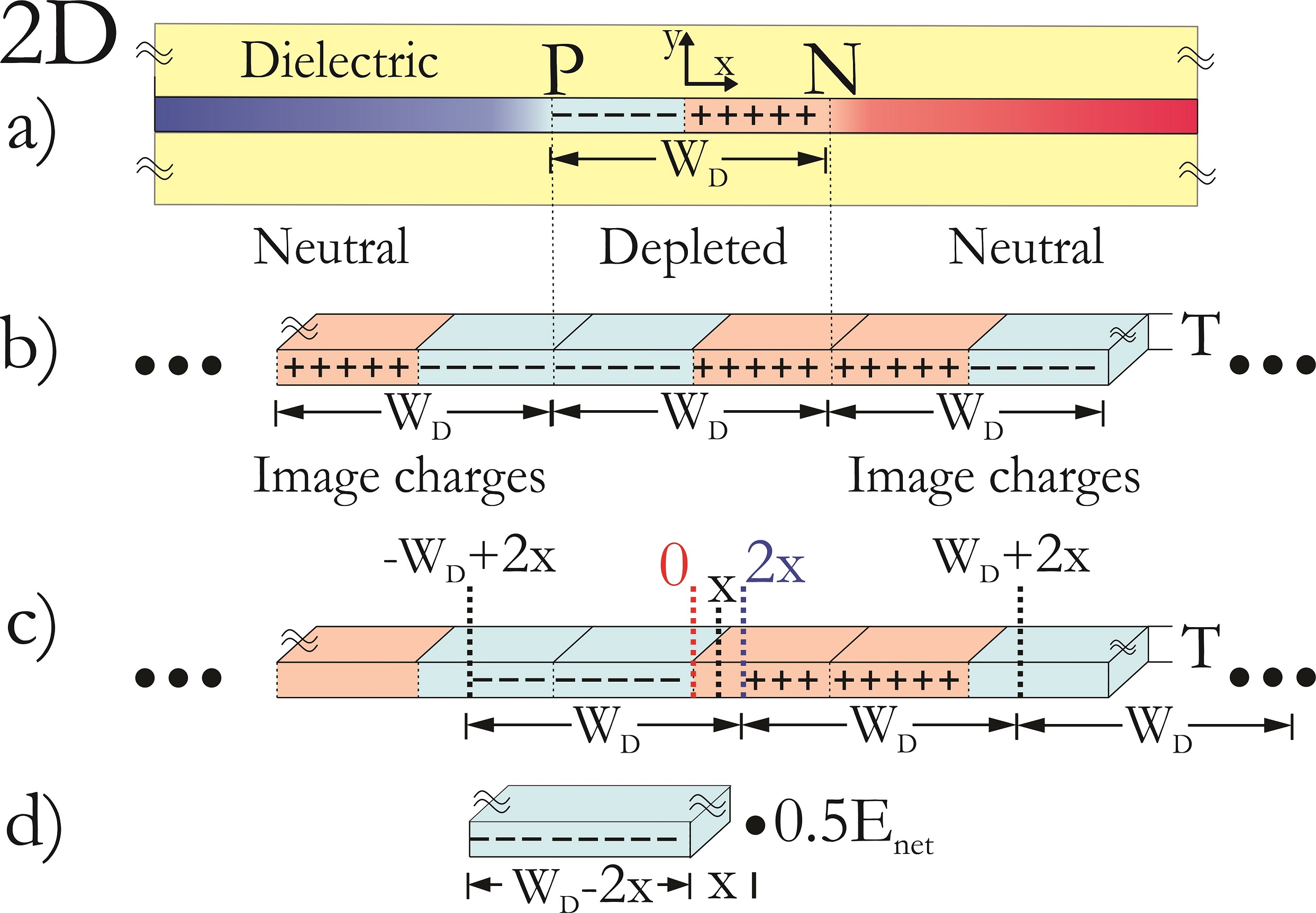}
                \label{fig:Same_EoT}
        \end{subfigure}%
        \caption{Illustration of the analysis method in a 2D PN junction similar to Fig. 2.}\label{fig:Fig1}
\end{figure}
\begin{figure}[!b]
        \centering
        \begin{subfigure}[b]{0.27\textwidth}
               \includegraphics[width=\textwidth]{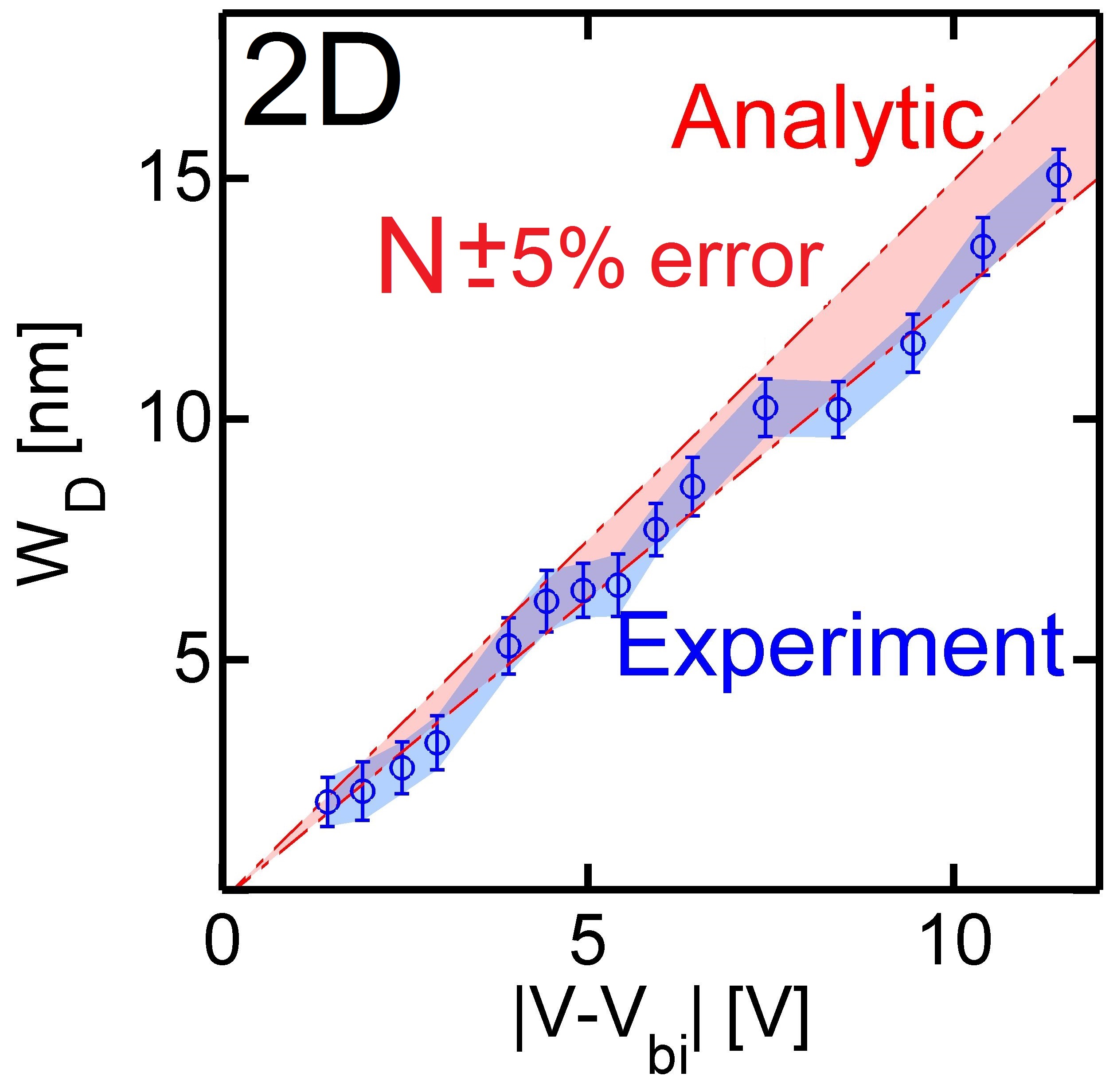}
                \label{fig:Same_EoT}
        \end{subfigure}%
        \vspace{-1.5\baselineskip}                    
        \caption{Comparison between analytical and experimental results of a 2D junction with $\pm 5 \%$ tolerance in measured N.}\label{fig:Fig1}
\end{figure}    
\begin{multline}
E_{net}^{2D} =   \frac{-E_0}{2} ln\left( \frac{4(W_D-x)^2 + T^2}{4x^2 + T^2} \right) \\
+ 2 E_0 \left( x~atan\left(\frac{T}{2x}\right) - (x - W_D) ~ atan\left(\frac{T}{2(x - W_D)}\right) \right)
\label{eq:opt1}
\end{multline}
where $E_0$ is defined as
\begin{equation} 
E_0 = \frac{q N}{\pi \epsilon}
\label{eq:opt1}
\end{equation}
To simplify the solution, the flake thickness can be assumed to be small compared with the depletion width ($T \ll W_D$ ) which leads to:
\begin{equation} 
E_{net}^{2D} \approx -E_0 ( ln(W_D - x) - ln(x) )
\label{eq:opt1}
\end{equation}
The potential is obtained from integrating electric field:
\begin{equation}
\label{eq:V2D}
\frac{V^{2D}(x)}{E_0} \approx  (x - W_D) ln(W_D -x)+W_D ln(W_D)-x~ln(x) 
\end{equation}
Evaluating (\ref{eq:V2D}) at $x$ = $W_D$/2 gives $W_D$ 
\begin{equation}
\label{eq:opt1}
W_D^{2D} \approx \frac{\Delta V}{ln(4) ~ E_0}
\end{equation}
Unlike in 3D, $W_D$ has a linear dependence on both $\Delta V$ and $\epsilon$ and is inversely proportional to $N$. To validate this significant difference from conventional PN junction behavior, the analytical results are compared with the experimental measurements of a PN junction made from 2DEGs (2 Dimensional Electron Gases). Fig 4 shows a good match between the analytic calculations and experimental data.  
\begin{figure}[!t]
        \centering
        \begin{subfigure}[b]{0.4\textwidth}
               \includegraphics[width=\textwidth]{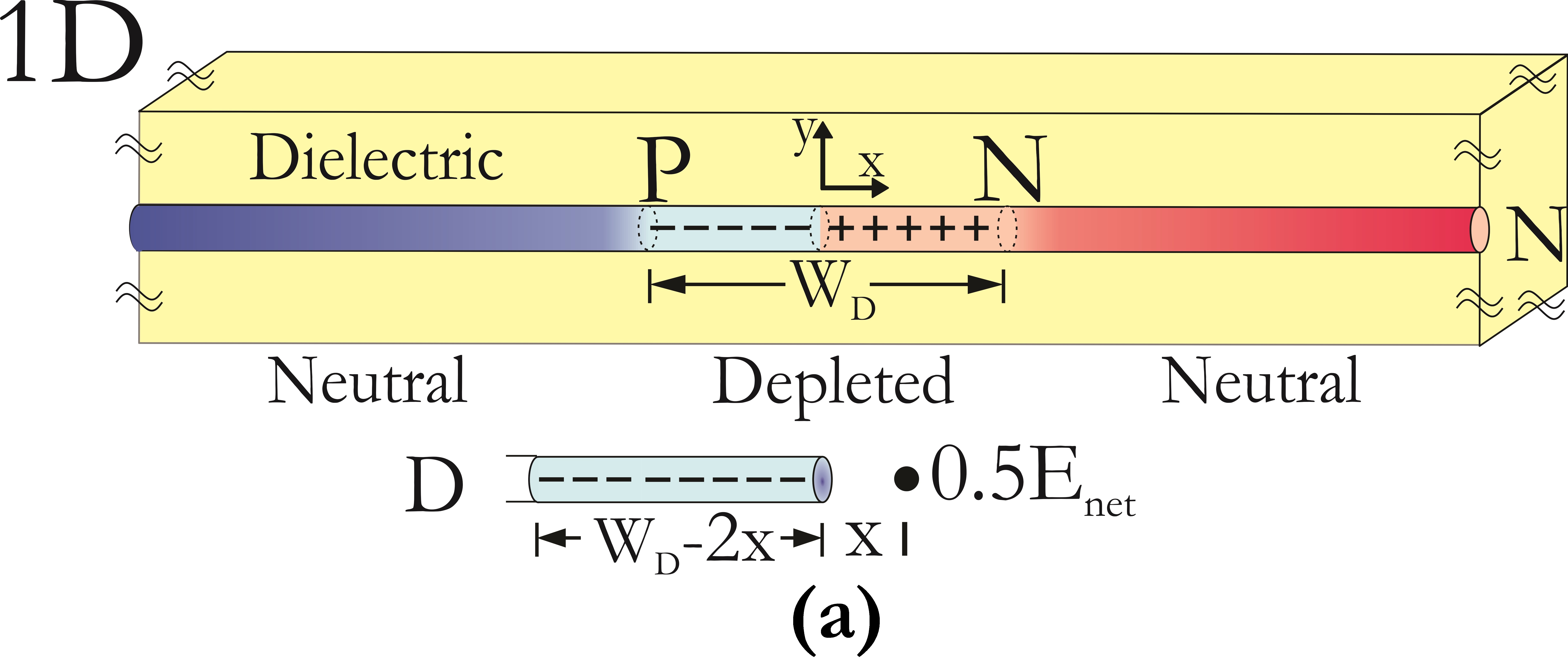} 
                \vspace{-0.8\baselineskip}                
                \label{fig:Same_EoT}
        \end{subfigure}%
        \quad
        \begin{subfigure}[b]{0.4\textwidth}
               \includegraphics[width=\textwidth]{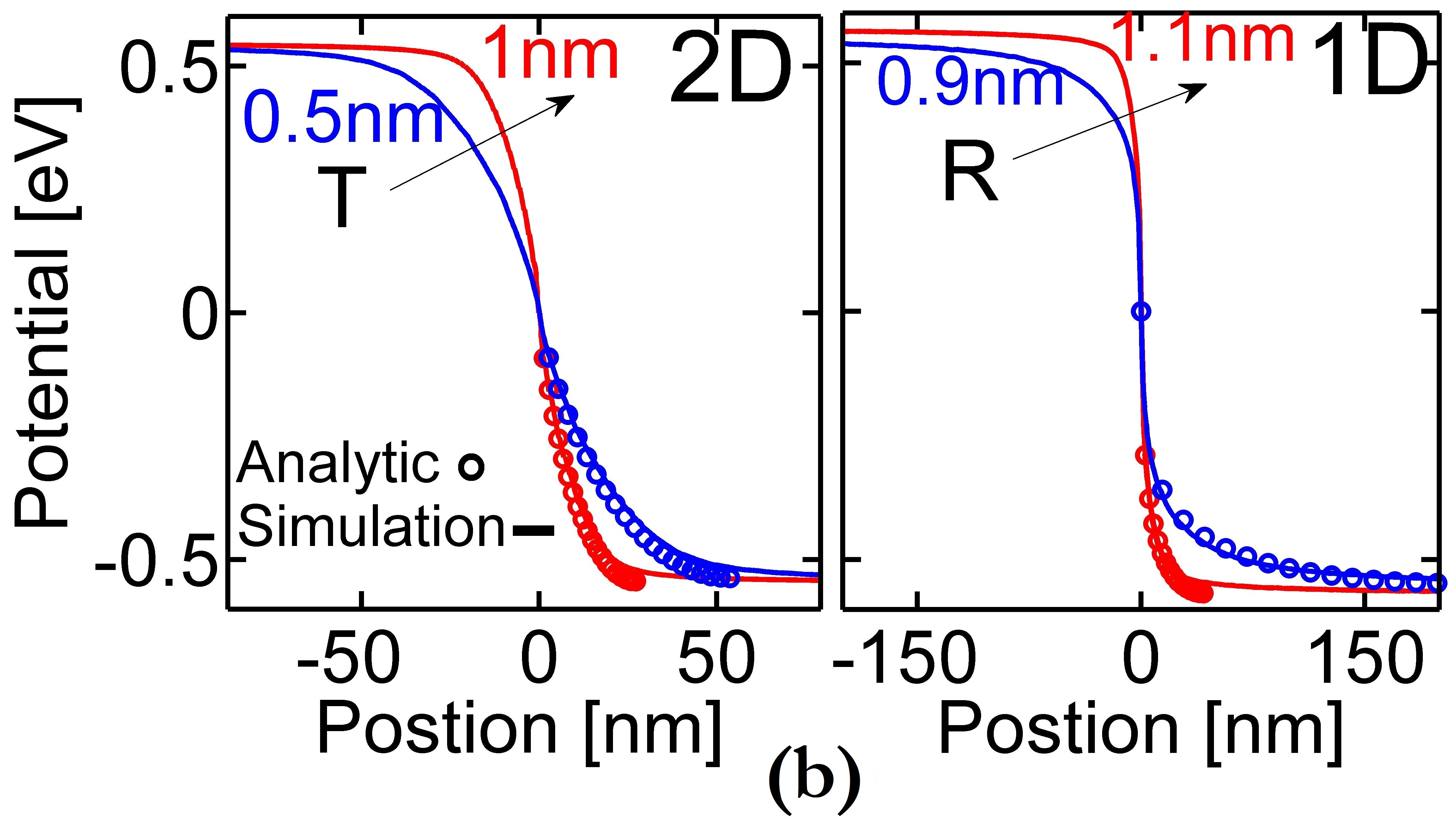} 
               \vspace{-1.5\baselineskip}
                \label{fig:Same_EoT}
        \end{subfigure}%
        \caption{a) Structure and uncompensated charges contributing to $E_{net}$ in 1D PN junction. b) Simulated (lines) and analytic (circles) potential profile of 2D (left) and 1D (right) junctions with different thicknesses. }\label{fig:Fig1}
\end{figure}
The analysis of 1D PN junction is similar to that of 2D. The only difference is that $E_{net}$ is due to a cylinder of charge with a length of $W_D - 2x$ and diameter $D$ at a distance $x$ (Fig. 5a):
\begin{equation}
E_{net}^{1D} =  -\frac{q N}{\epsilon}  \left( (W_D - 2x) - L_{W_D - x} + L_x \right)
\label{eq:opt1}
\end{equation}
where the function $L_x$ is defined as $\sqrt{R^2 + x^2}$.
and $R$ is the radius of nanowire or nanotube. The potential is obtained to be:
\begin{multline}
\label{eq:opt1}
\frac{V^{1D}(x)}{V_0^{1D}} = sinh^{-1}(\frac{x}{R}) + sinh^{-1}(\frac{W_D-x}{R}) - sinh^{-1}(\frac{W_D}{R}) \\
+ \frac{ x L_x + (W_D-x) L_{W_D-x} - 2x(W_D-x) }{R^2} 
\end{multline}
where $V_0^{1D}$ is defined as $\frac{q N}{2 \epsilon} R^2$.
The depletion widths of 1D PN junctions exhibit an exponential dependence on $\Delta V \epsilon / N$
\begin{equation}
\label{eq:opt1}
W_D^{1D} \approx R ~ exp \left(\frac{2 \epsilon \Delta V}{q N R^2} - \frac{1}{2}\right)
\end{equation}
	
Figure 5b compares the results of analytical model and numerical simulation of 2D and 1D PN junctions with different thicknesses showing a good agreement. Moreover, it demonstrates the significant impact of the flake thickness and nanowire diameter on the potential profile. Increasing the thickness of low dimensional material enhances the depleted charge and electric field considering a constant volume doping density. 
\begin{table}[b]
               \vspace{+0.2\baselineskip}
\normalsize
\captionof{table}{V(x) in 2D and 3D valid for all thicknesses.}
               \vspace{-0.4\baselineskip}
\resizebox{0.5\textwidth}{!}{
\renewcommand{\arraystretch}{1.15}
\centering
\begin{tabular} { |c|c| }
  \hline
  V(x) & $ \frac{q N}{2 \epsilon} \left( \Gamma(W_D) - \Gamma(x) - \Gamma(W_D-x) \right) $ \\ \hline  
  2D & $ \Gamma(x) = \frac{1}{\pi} \beta ~ atan(\frac{2x}{T}) ( \frac{3}{4 }T^2 - x^2 ) +  \frac{\beta}{2} x^2 +  \frac{1}{\pi} x T ln(\frac{T^2}{4} + x^2)$   \\ \hline
  1D & $ \Gamma(x) = -R^2 asinh( \frac{x}{R}) - \beta ~x \sqrt{x^2 + R^2} + \frac{\beta x}{2}(x- \frac{W_D}{2}) $ \\ \hline 
$\beta$ & $\beta_{2D} = \frac{3}{2} - \frac{1}{\pi} atan(ln(\frac{2T}{W_D}))$ and $\beta_{1D} = \frac{3}{2} - \frac{1}{\pi} atan(ln(\frac{2 R}{W_D}))$  \\ \hline 
\end{tabular}}
\label{Table}
\end{table}

To make the analytic model work for thicknesses from 0 to $\infty$, the exact equations must be used. It is noteworthy that the potential attains an additional factor $\beta$ for large thicknesses due to the impact of distant charges as shown in table II. 
Figure 6 shows the depletion width as a function of junction thickness in different dimensions. For a thickness beyond depletion width, the fringing field gets screened at the surface and $W_D$ gets close to the corresponding value in 3D. However, for smaller thicknesses, $W_D$ deviates from 3D case. Especially, 1D case shows a significantly higher $W_D$ values and increased sensitivity to the diameter. $W_D$ has a similar response to variations in $N$. The increased sensitivity of $W_D$ with respect to junction parameters in low dimensions enables the possibility of new sensors. 

In summary, using a new analysis, the electrostatics of PN junctions in low dimensional material systems have been shown to differ significantly from the 3D junctions. Reducing the dimensionality increases the depletion width and its sensitivity to doping and thickness of PN junction, attractive towards sensing applications. The analytic results match closely with experimental measurements and numerical simulations.  
\begin{figure}[!t]
        \centering
        \begin{subfigure}[b]{0.45\textwidth}
               \includegraphics[width=\textwidth]{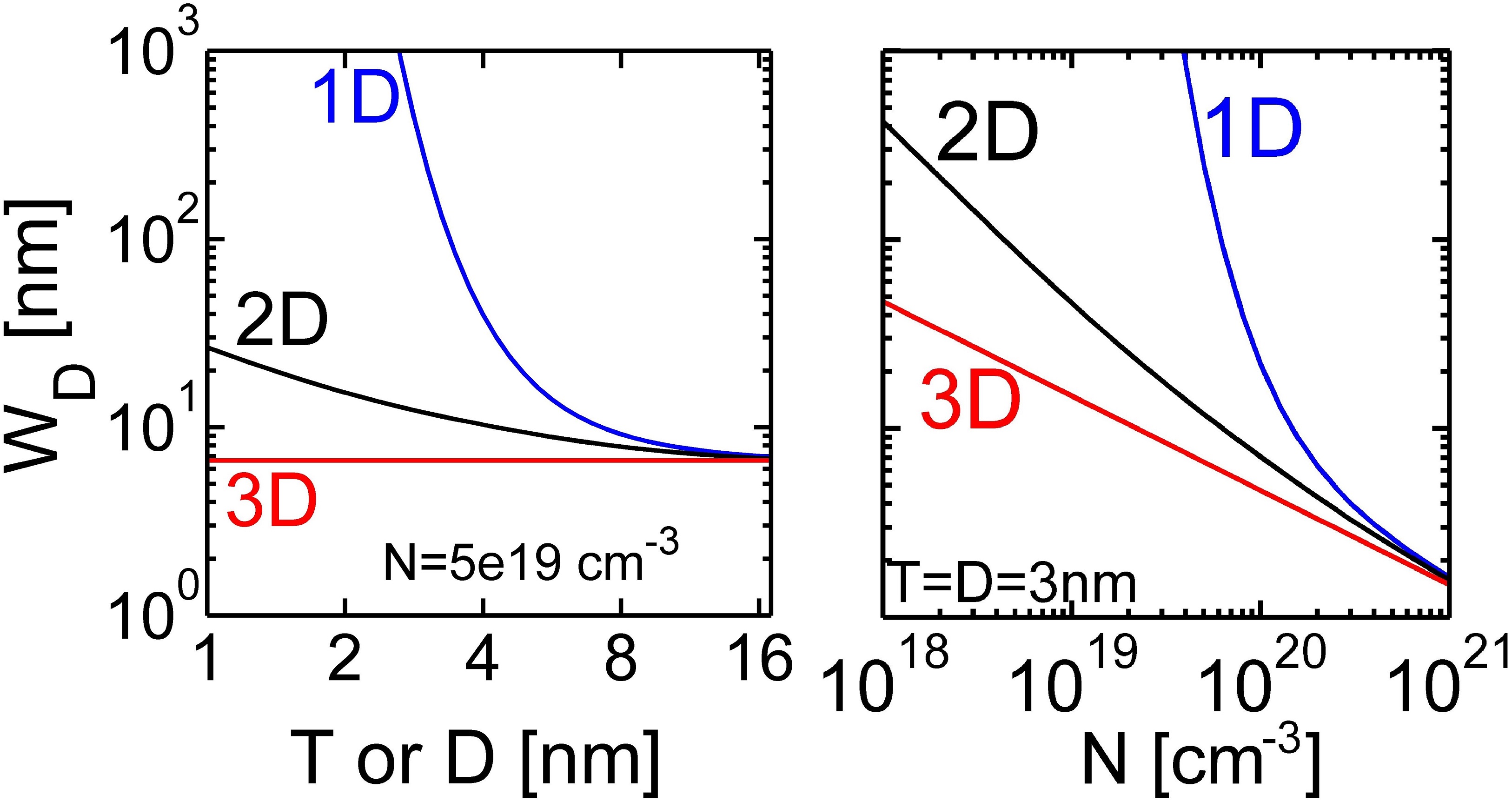}
               \vspace{+.3\baselineskip}
                \label{fig:Same_EoT}
        \end{subfigure}%
        \vspace{-1.5\baselineskip}        
        \caption{$W_D$ versus thickness (left) and N (right). }\label{fig:Fig1}
\end{figure}
\clearpage
\newpage

\bibliographystyle{ieeetr}
\bibliography{thesis}

\end{document}